\documentclass[12pt]{article}
\usepackage[T1]{fontenc}
\usepackage[utf8]{inputenc}
\usepackage[margin=1in]{geometry}
\usepackage[affil-it]{authblk} 
\usepackage{etoolbox}
\usepackage{lmodern}
\usepackage{cite}
\usepackage{changes}

\usepackage{amssymb}

\usepackage{float}
\DeclareUnicodeCharacter{030C}{\v{c}}

\title{Quantum photonics in triangular-cross-section nanodevices in silicon carbide}

\def\correspondingauthor{\footnote{Corresponding author: smajety@ucdavis.edu}}

\author[1]{Sridhar Majety \correspondingauthor{}}
\author[2]{Victoria A. Norman}
\author[1]{Liang Li}
\author[3]{Miranda Bell}
\author[1]{Pranta Saha}
\author[1]{Marina Radulaski}
\affil[1]{Department of Electrical and Computer Engineering, University of California, Davis, CA 95616, USA}
\affil[2]{Department of Physics, University of California, Davis, CA 95616, USA}
\affil[3]{Department of Materials Science and Engineering, University of California, Davis, CA 95616, USA}

\date{\vspace{-2em}} 

\usepackage{graphicx}
\usepackage{amsmath}
\usepackage{braket}
\usepackage{titlesec}
\usepackage{siunitx} 
\begin{document}

\maketitle
\begin{abstract}
    Silicon carbide is evolving as a prominent solid-state platform for the realization of quantum information processing hardware. Angle-etched nanodevices are emerging as a solution to photonic integration in bulk substrates where color centers are best defined. We model triangular cross-section waveguides and photonic crystal cavities using Finite-Difference Time-Domain and Finite-Difference Eigensolver approaches. We analyze optimal color center positioning within the modes of these devices and provide estimates on achievable Purcell enhancement in nanocavities with applications in quantum communications. Using open quantum system modeling, we explore emitter-cavity interactions of multiple non-identical color centers coupled to both a single cavity and a photonic crystal molecule in SiC. We observe polariton and subradiant state formation in the cavity-protected regime of cavity quantum electrodynamics applicable in quantum simulation. 
\end{abstract}

\section{Introduction}

Color center photonics has been on the rise in explorations of quantum information processing. Initially, the negatively charged nitrogen vacancy in diamond was explored as a solid-state qubit that can be individually initialized, manipulated and measured with high fidelity even at room temperature \cite{neumann2008multipartite, norman2020novel}. Owing to the lack of availability of wafer-scale substrates and color center emission at telecommunications wavelengths, new material hosts have been explored as potential defect hosts. Among them, silicon carbide (SiC) gained attention as a substrate whose color centers have long spin coherence times \cite{widmann2015coherent, christle2015isolated, seo2016quantum}, excellent brightness \cite{castelletto2014silicon} and manipulations of nuclear spin \cite{falk2015optical, klimov2015quantum}, and are suitable for applications in quantum computing \cite{kok2007linear, weber2010quantum, klimov2015quantum}, quantum photonics \cite{atature2018material, koehl2011room}, quantum communications \cite{hensen2015loophole, castelletto2014silicon,wang2018bright,falk2013polytype}, quantum sensing \cite{mamin2013nanoscale, kucsko2013nanometre, degen2017quantum, simin2016all, falk2014electrically}, quantum metrology \cite{giovannetti2011advances, wolfowicz2018electrometry} and quantum simulation \cite{cai2013large, wang2015quantum, aspuru2012photonic, georgescu2014quantum, Harris2017}. In addition, SiC has benefits like large bandgap, high thermal conductivity, strong second-order non-linearity, infra-red emitting color centers, and has a decades-long industrial presence. 

It has been well studied in atomic \cite{kimble1998strong} and quantum dot \cite{reithmaier2004strong, buckley2012engineered} systems that optical resonators enhance emitter properties, from Purcell emission enhancement and coupling to a desired mode to polariton physics and cavity Quantum Electrodynamics (cQED). These effects have been used in high-speed quantum light generation, all-optical switching, creation of hybridized light-matter states, and more. The advantages of integrated photonics with color centers lie in scalability granted by the combination of the solid-state platform and the nearly identical optical properties of ensembles \cite{radulaski2017nonclassical}, inaccessible to the previously studied systems.

A key requirement for nanophotonic integration is the ability to grow or place high quality, high refractive index thin films on a contrasting substrate which can facilitate high optical confinement in nanostructures. For example, the ability to bond silicon on insulator in a scalable way launched silicon as a strong material platform for photonics applications \cite{ye2013review}. Although commercial scale high quality crystalline SiC wafers grown using homoepitaxy (MBE, CVD) or sublimation have transferred into color center research, for example by \emph{in situ} doping of vanadium centers in 4H-SiC \cite{karhu2019cvd}, this approach does not provide an undercutting layer needed to generate freestanding photonic devices. On the other hand, growth of 3C-SiC on silicon has been established in MEMS research \cite{ferro20153c}, but has not lent itself well for color center formation in nanoscale films due to the lattice mismatch damage \cite{calusine2014silicon}. These substrate limitations have been circumvented via photo-electrochemical etching \cite{bracher2017selective} and SiC thinning and bonding to an insulator \cite{lukin20204h}. While these approaches successfully integrated and manipulated color centers in a nanocavity, the fabrication processes have been limited in terms of either type or size of the substrate, requiring a specific doping profile or a limited chip area.

Another bulk substrate nanofabrication approach, initially championed in diamond \cite{burek2012free}, has been emerging in silicon carbide: the angle-etching method \cite{song2018high}. In contrast to the more traditional (rectangular) profile, the photonic devices created via angle-etch have a triangular cross-section. Such SiC processing methods are currently based on the Faraday-cage assisted etching; still, wafer scale processes could be implemented using ion beam etching \cite{atikian2017freestanding}. Triangular device geometry resolves the problem of the unavailability of heteroepitaxial thin film color center substrates because it can be implemented in bulk substrates of any polytype or doping level. However, due to the novelty of the approach, not much is known about the optical properties and parameters related to the color center integration into angle-etched SiC devices.

This paper studies triangularly shaped nanophotonic devices in SiC suitable for the development of quantum communication and quantum simulation hardware. Nitrogen vacancy and vanadium impurity in 4H-SiC are of particular interest in this work, since their emission wavelengths are in the telecommunications  range which makes them useful candidates for long-distance quantum information distribution. Using Lumerical MODE and FDTD software packages, we examine the TE and TM polarized modes in the near infrared (NIR) part of the spectrum supported in triangular waveguides with variable etch-angles, characterizing their effective index of refraction and the depth of the mode maximum. We design high $Q$/$V$ ratio photonic crystal cavities and analyze their resonance parameters and potential Purcell enhancement for variable triangular profiles. We extend this design into a photonic crystal molecule with coupled cavity array integration in sight. Finally, using open quantum system modeling, we explore emitter-cavity interactions of multiple non-identical color centers coupled to both a single cavity and a photonic crystal molecule, observing polariton and subradiant states formation.

\section{SiC color center emission properties}

While all SiC polytypes can host color centers in highly crystalline form, their processing into a photonic platform can induce irreversible lattice damage that prevents formation of well-defined quantum emitters. Advanced 4H-SiC carbide processing techniques have made this polytype a promising candidate for color-center integrated nanophotonics. 4H-SiC has a hexagonal close-packed lattice with two inequivalent lattice sites: the hexagonal site, $h$, and the quasi-cubic site, $k$; this inequivalency allows for 4 defect orientations ($hh$, $hk$, $kh$, and $kk$) in a two-site defect, with each defect orientation contributing its own zero-phonon line (ZPL) to the spectrum. The most prominently studied centers in 4H-SiC polytype are shown in Table \ref{tab:Table1}. As solid-state emitters, color centers interact with the host lattice which modifies their optical properties. The proportion of radiative emission occurring through the zero-phonon transition compared to the transitions to all vibronic levels is given by the Debye-Waller (DW) factor $\xi$. 
Integration into high quality factor ($Q$) and small mode volume ($V$) cavities can rebalance this ratio via the Purcell effect. The radiative emission enhancement is given by the Purcell factor:
\[
    F_P = \left[\frac{3}{4\pi^2} \bigg(\frac{\lambda}{n} \bigg)^{3} \bigg(\frac{Q}{V} \bigg) \right]\left[ \left| \frac{E}{E_{\textnormal{max}}} \right| \cos(\phi) \right] \xi
\]
    The terms $\left| E/E_{\textnormal{max}} \right| \cos(\phi)$ and $\xi$ refer to the cavity-color center spatial overlap and spectral matching respectively. A photonic crystal cavity (PCC) enabling large Purcell enhancement ensures a high emission rate of indistinguishable photons needed for the implementation of quantum entanglement schemes \cite{soykal2016silicon}.

\begin{table}[!htb]
\centering
\resizebox{\columnwidth}{!}{\begin{tabular}{c c c c c c}
\hline

Defect&ZPL wavelength&DW factor& Emission&Calculated & Purcell\\
&(nm)&&lifetime &$g_{max}/2\pi $ (GHz) & enhancement\\
\hline
V1' (V$_{\textnormal{Si}}^-$) \cite{nagy2018quantum}&858&19\%&(5.6 $\pm$ 1.2) ns & 3.41 & 120 $\pm$ 10 \cite{lukin20204h}\\
V1 (V$_{\textnormal{Si}}^-$) \cite{PhysRevApplied.13.054017}&861&8\%&(5.5 $\pm$ 1.4) ns & 2.23 & 25 \cite{bracher2017selective}\\
V2 (V$_{\textnormal{Si}}^-$) \cite{PhysRevApplied.13.054017,hain2014excitation}&918&9\%&6 ns &2.27 & -\\
Cr$^{4+}$ \cite{koehl2017resonant}&1070&75\% &145.8 $\pm$ 9.7 $\mu$s &0.04 & -\\
V$_{\textnormal{C}}$V$_{\textnormal{Si}}$$^0$ ($kk$) \cite{christle2017isolated,falk2014electrically}&1130&(5.3 $\pm$ 1.1)\%&(14 $\pm$ 3) ns & 1.14 & 53\cite{crook2020purcell}\\
V$_{\textnormal{Si}}$N$_{C}$  \cite{doi:10.1021/acs.nanolett.0c02342} & 1176 - 1243 & - & 2.2 ns & -\\
V$^{4+}$ ($\alpha$) \cite{wolfowicz2020vanadium}&1279&25\%&167 ns & 0.72 & -\\
V$^{4+}$ ($\beta$) \cite{wolfowicz2020vanadium}&1335&50\%&41 ns & 2.04 & ~$10^5$\\

\hline
\end{tabular}}
\caption{Emission parameters for various color centers in 4H-SiC. V$_{\textnormal{Si}}^-$, V$_{\textnormal{C}}$V$_{\textnormal{Si}}$$^0$, V$_{\textnormal{Si}}$N$_{C}$, Cr$^{4+}$, and V$^{4+}$ stand for silicon vacancy, divacancy, nitrogen vacancy,  chromium impurity, and vanadium impurity respectively. V1 and V1' are ZPLs of hexagonal site ($h$) silicon vacancy, V2 is the ZPL of cubic site ($k$) silicon vacancy. A DW factor for V$_{\textnormal{Si}}$N$_{\textnormal{C}}$ has not been reported in literature and the upper and lower values of the ZPL range come from the hk and kh orientations respectively. V$^{4+} (\alpha)$ (V$^{4+} (\beta)$) likely corresponds to the $h$ ($k$) site \cite{wolfowicz2020vanadium}. The temperature associated with each DW factor measurement is 300 K, except for Cr$^{4+}$ which was measured at 30 K. The Purcell enhancement measurements were performed in fabricated devices except for the V$^{4+}(\beta)$ which is the result of this work.}
\label{tab:Table1}
    
\end{table}

\section{Triangular cross-section SiC waveguide}

\begin{figure}[h!]
    \centering
    \includegraphics[scale=0.85]{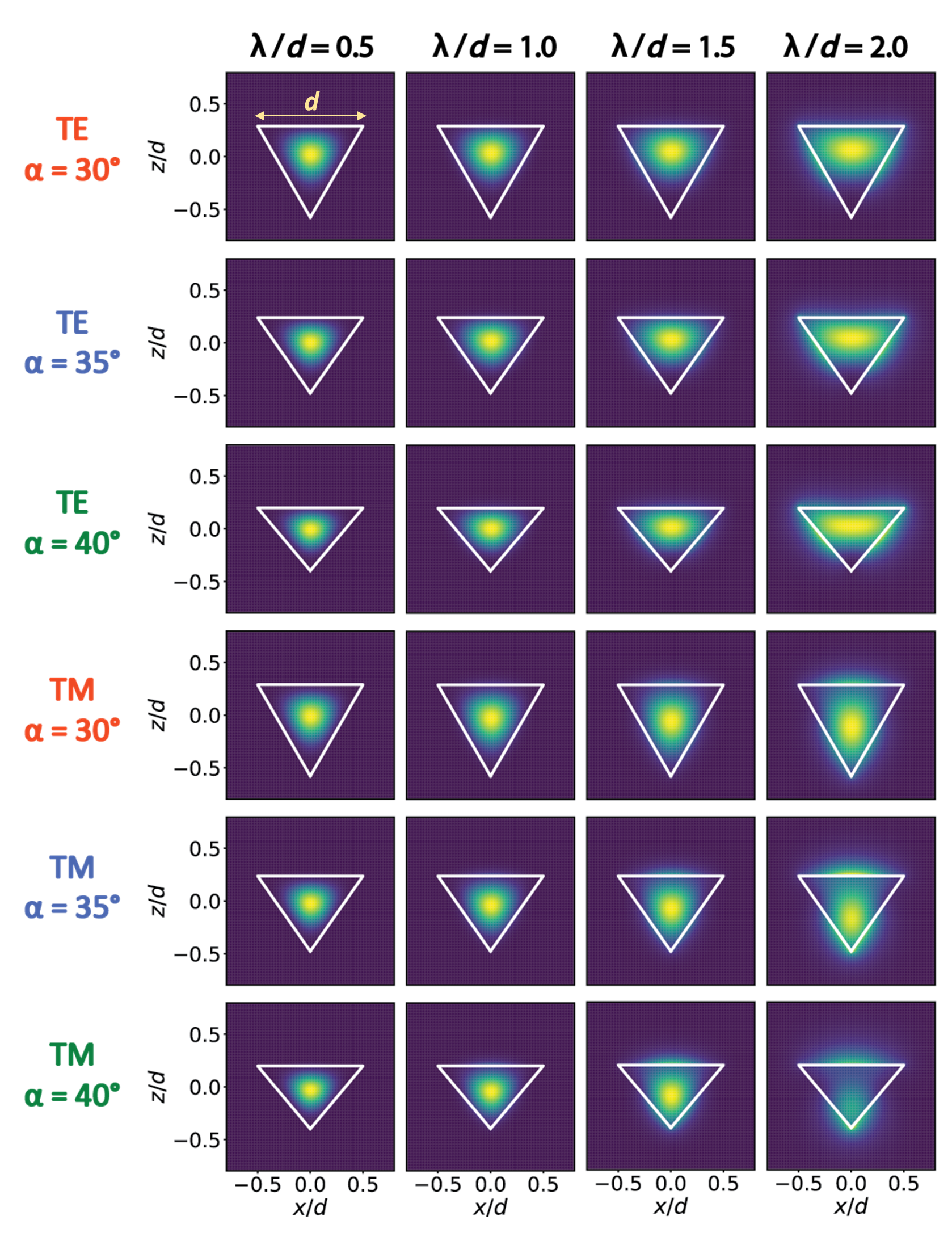}
    \caption{Electric field intensity profiles $|E|^2$ of TE and TM polarized modes supported by triangular-cross-section SiC waveguides with $0.5 \leq \lambda/d \leq 2$ and $30^\circ \leq \alpha \leq 40^\circ$. A mesh dimension of 10 nm was used in the x, y, z directions and a constant width $d$ = 0.754 $\mu$m was used. Color bar is the same as in Figure \ref{fig:fig4}. 
    }
    \label{fig:fig2}
\end{figure}

A waveguide uses total internal reflection to confine propagating light in the high refractive index material. The triangular cross-section waveguide modeled in this study using Finite-Difference Eigensolver (Lumerical MODE) was parametrized by the wavelength to the top-width ratio $\lambda$/$d$ and the half-angle $\alpha$. The fundamental TE and TM modes were studied for $0.5 \leq \lambda/d \leq 2$ and $30^\circ \leq \alpha \leq 40^\circ$ with obtained mode profiles shown in Figure \ref{fig:fig2}.

\begin{figure}[htbp]
    \centering
    \includegraphics[width=\linewidth]{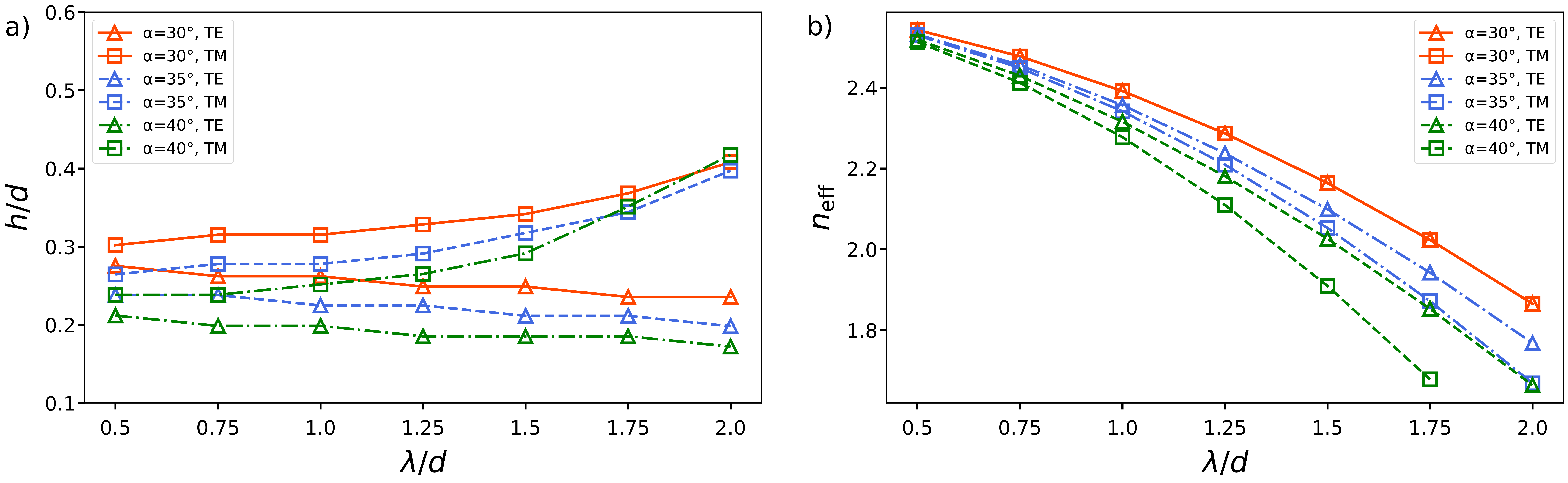}
    \caption{Normalized depth $h/d$ of the electric field intensity maximum $|E|^2_{\textnormal{max}}$ and the effective index of refraction of TE and TM modes supported in a triangular-cross-section SiC waveguide, as a function of the geometry parameters.    
    }
    \label{fig:fig1}
\end{figure}

To find the optimal positioning of the color center in the waveguide, we analyzed the depth $h$ at which the electric field intensity maximum ($|E|^2_{\textnormal{max}}$) of a mode is positioned relative to the waveguide's top plane. As $\lambda$/$d$ increases, $|E|^2_{\textnormal{max}}$ for the fundamental TE (TM) polarized mode shifts towards (away from) the top waveguide surface i.e., the depth $h$ decreases (increases). This behavior is consistent for all studied half-angle values $\alpha$, as shown in Figure \ref{fig:fig1}a. For higher values of $\lambda$/$d$, the mode becomes less confined and the evanescent losses into the surrounding medium dominate, as can be observed in mode profiles in Figure \ref{fig:fig2}. These observations are consistent with the trend in the effective refractive index ($n_{\textnormal{eff}}$) values shown in Figure \ref{fig:fig1}b. The effective refractive index for a mode is given by the ratio of propagation constant of that mode ($\beta$) to the wavenumber in vacuum ($k_0$). When $n_{\textnormal{eff}} \sim 1.5$, significantly below the SiC refractive index $n = 2.6$, $|E|^2_{\textnormal{max}}$ for the TM mode lies outside the waveguide and results in a lossy propagation. These results give an understanding into the range of half-angle and $\lambda/d$ values suitable for fabricating triangular cross-section waveguides that are able to couple the fundamental TE/TM modes with other photonic elements.

\section{Triangular cross-section SiC photonic crystal cavity}

\begin{figure}[h!]
    \centering
    \includegraphics[scale=0.75]{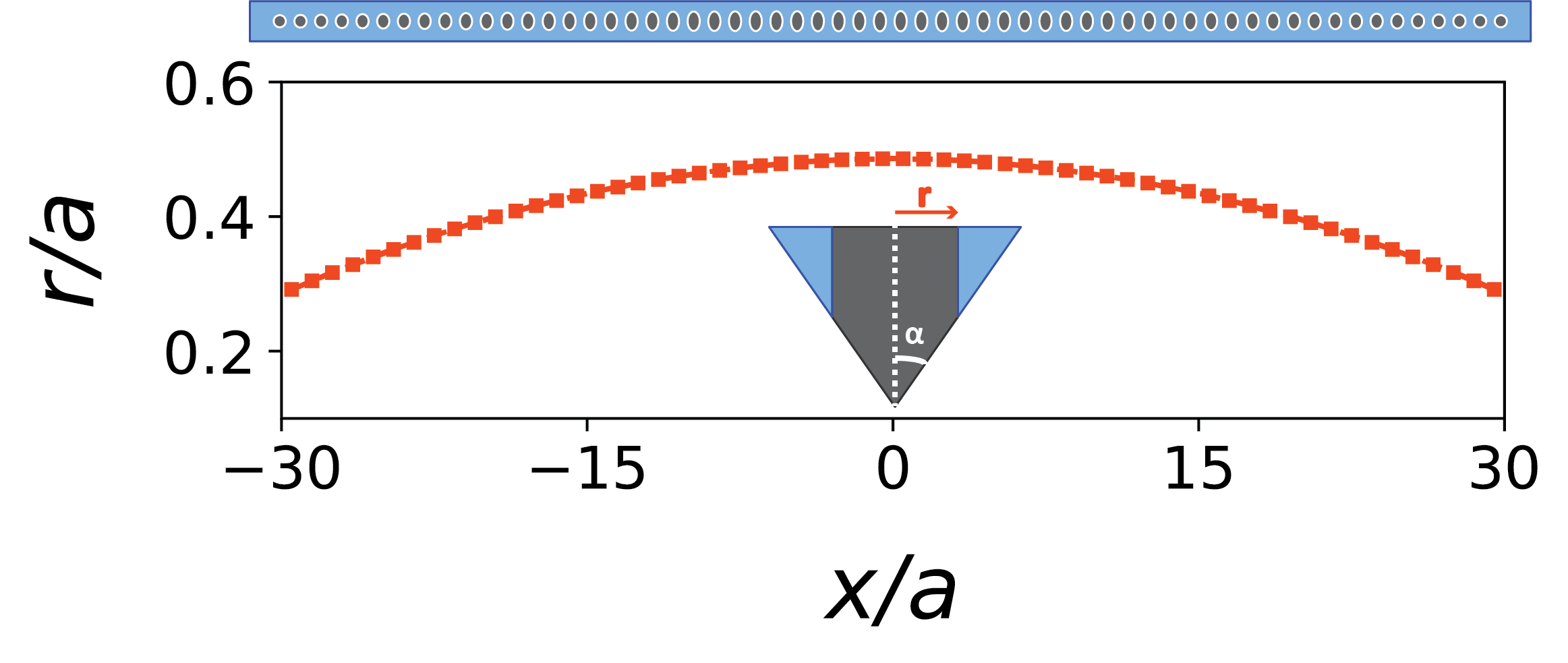}
    \caption{Triangular nanobeam cavity structure and the parabolic variation of its hole's major radius $r$ along the length of the beam.
    }
    \label{fig:fig3}
\end{figure}

Photonic crystal cavities are nanostructures with periodic variation of refractive index which localize photons to sub-wavelength mode volumes. The triangular nanobeam photonic crystal cavities explored in this study have elliptical air holes whose major axis radius is tapered along the device, which have proved successful in diamond color center research in achieving ultra-high Q-factors and ultra-small mode volumes \cite{quan2011deterministic, burek2014high} and more importantly, offer better mode confinement and scalability than the well-explored methods. As shown in Figure \ref{fig:fig3}, the cavity has a width $d = 1.94a$ and consists of 30 through holes on either side of the center of the cavity with uniform spacing between the holes, lattice constant $a = 387.8$ nm. The minor radius of the holes is $0.292a$ and to minimize scattering and maximize Q-factor, the major radius is quadratically varied from $r{_1}= 0.486a$ to $r{_{30}} = 0.292a$ over 30 holes. The beam has a triangular cross-section with a half-angle $\alpha$ = \ang{35}.

\begin{figure}[h!]
    \centering
    \includegraphics[scale=1]{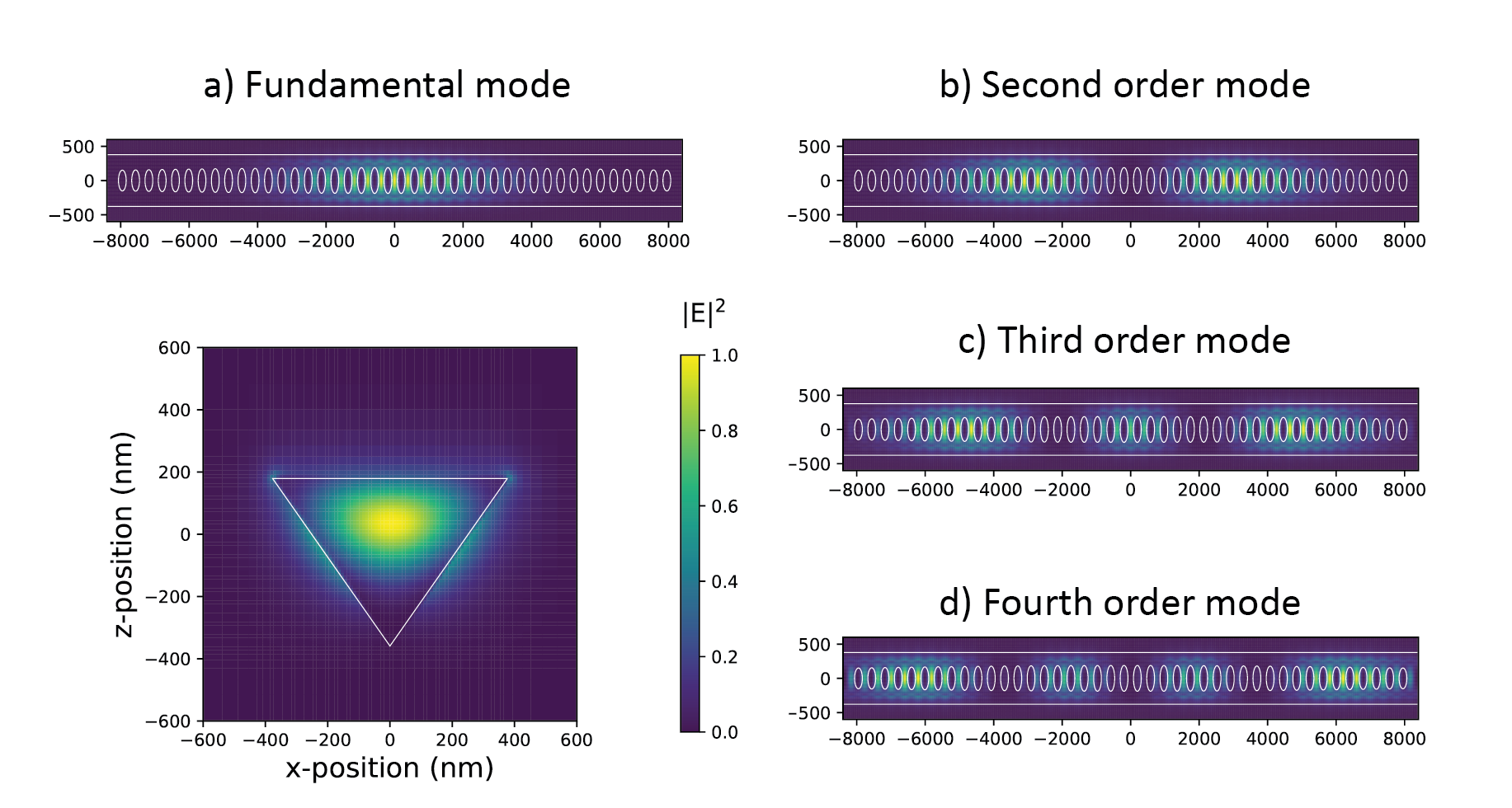}
    \caption{a) The top and the cross-sectional profile of the fundamental resonant mode. b-d) Top-view profile of the first three higher order modes in the cavity.}
    \label{fig:fig4}
\end{figure}

\begin{figure}[h!]
    \centering
    \includegraphics[width=\linewidth]{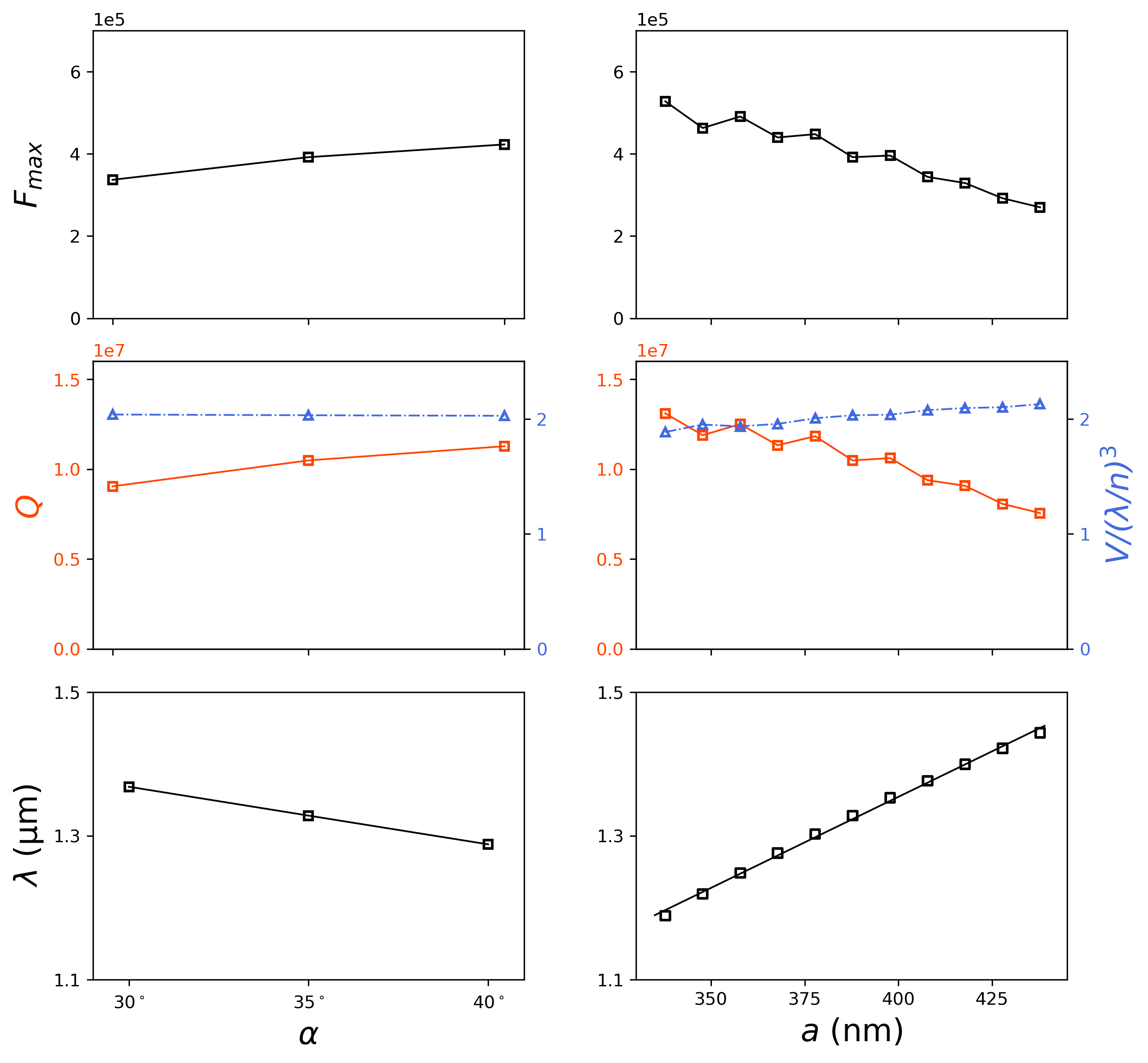}
    \caption{Dependence of the resonant wavelength $\lambda$, the quality factor $Q$, the mode volume $V/(\lambda/n)^3$ and the maximum Purcell enhancement $F_{\textnormal{max}}$ on the half-angle $\alpha$ of the triangular cross-section and the lattice constant in the triangular nanobeam SiC cavity.
    }
    \label{fig:fig5}
\end{figure}

\begin{figure}[h!]
    \centering
    \includegraphics[scale=0.85]{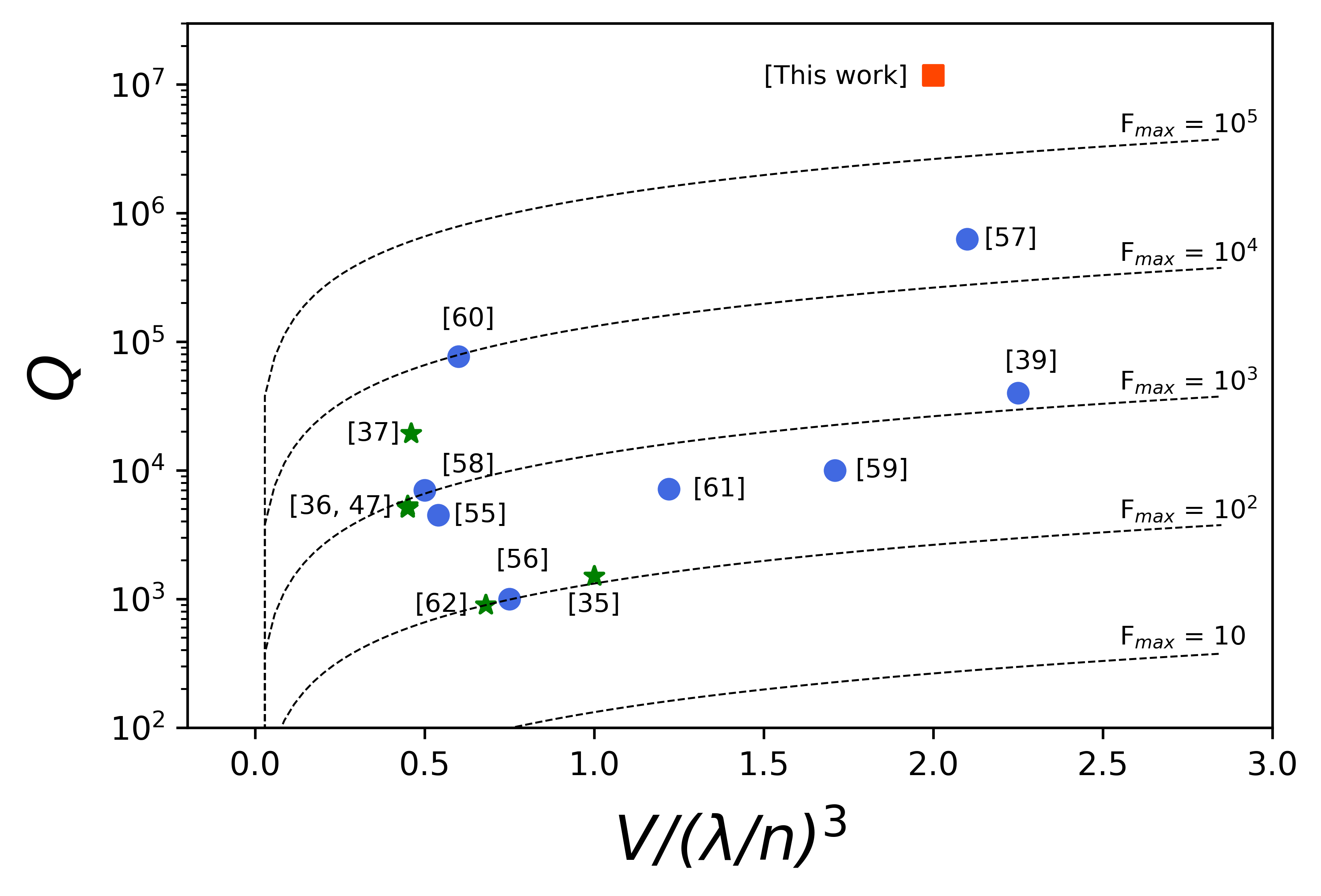}
    \caption{Comparison of experimentally demonstrated $Q$ and $V$ values of photonic crystal cavities in SiC - without color centers (blue circles) \cite{song2011demonstration,radulaski2013photonic, song2019ultrahigh, bracher2015fabrication, song2018high, yamada2012suppression,lee2015high, chatzopoulos2019high}, with integrated color centers (green stars) \cite{lukin20204h, calusine2014silicon, bracher2017selective, crook2020purcell, calusine2016cavity} and simulated value at 1300 nm in this work (red square). The dashed black lines indicate the different regions of maximally achievable Purcell factor values.
    }
    \label{fig:fig6new}
\end{figure}

The optical response of the triangular nanobeams was modeled using the Finite-Difference Time-Domain method in the Lumerical FDTD software package. In the simulations, SiC was defined as a dielectric with a constant refractive index $n= 2.6$. The electric field profile of the fundamental mode is shown in Figure \ref{fig:fig4}a. The fundamental cavity resonance occurs at $\lambda = 1324$ nm with a quality factor of $Q \sim 1.05 \times 10^7$ and mode volume of $V \sim 2(\lambda/n)^3$. Higher order TE-polarized longitudinal modes, useful for nonlinear optics processes involving multiple modes such as frequency conversion \cite{buckley2014nonlinear, buckley2014second, lukin20204h, yamada2014second}, are also supported in the cavity, as shown in Figure \ref{fig:fig4}b-d, where the cross-sectional profile is representative of the profiles of the electrical field nodes in all modes. Cavities with half-angle $30^\circ \leq \alpha \leq 40^\circ$ were simulated while maintaining the lattice constant and radii of the holes. It was observed that the resonant wavelength of the fundamental mode decreased with increasing $\alpha$, whereas the Q-factor and mode volume remained almost the same, as shown in Figure \ref{fig:fig5}. The impact of a changing lattice constant was studied across a range of 100 nm, which tuned the resonant wavelength from 1200 nm to 1500 nm, as shown in Figure \ref{fig:fig5}. The Q-factor and mode volume of the fundamental resonance for varying values of $a$ remained nearly constant (less than a factor of 2 for $Q$ and within $10\%$ for $V$). Thus, for different technological implementations of the inclines in angled etching, this tapered-hole design of a photonic cavity can be used to integrate various SiC color centers in the telecommunications wavelength range with strong Purcell enhancement of $F_{\textnormal{max}}$ in the range of $2.7 \times 10^5$ - $5.3 \times 10^5$ when the spatial and spectral overlap terms are unity. Figure \ref{fig:fig6new} compares the simulated $Q$ and $V$ values in this work to other photonic crystal cavities fabricated in SiC. The cavities without emitters achieved quality factors up to 630,000 \cite{song2011demonstration,radulaski2013photonic, song2019ultrahigh, bracher2015fabrication, song2018high, yamada2012suppression,lee2015high, chatzopoulos2019high} and those with emitters achieved up to 19,300 with Purcell enhancement up to 120 \cite{lukin20204h, calusine2014silicon, bracher2017selective, crook2020purcell, calusine2016cavity}, showing that our design bridges the gap between color center platforms and high Purcell enhancement devices.

\begin{figure}[h!]
    \centering
    \includegraphics[width=\linewidth]{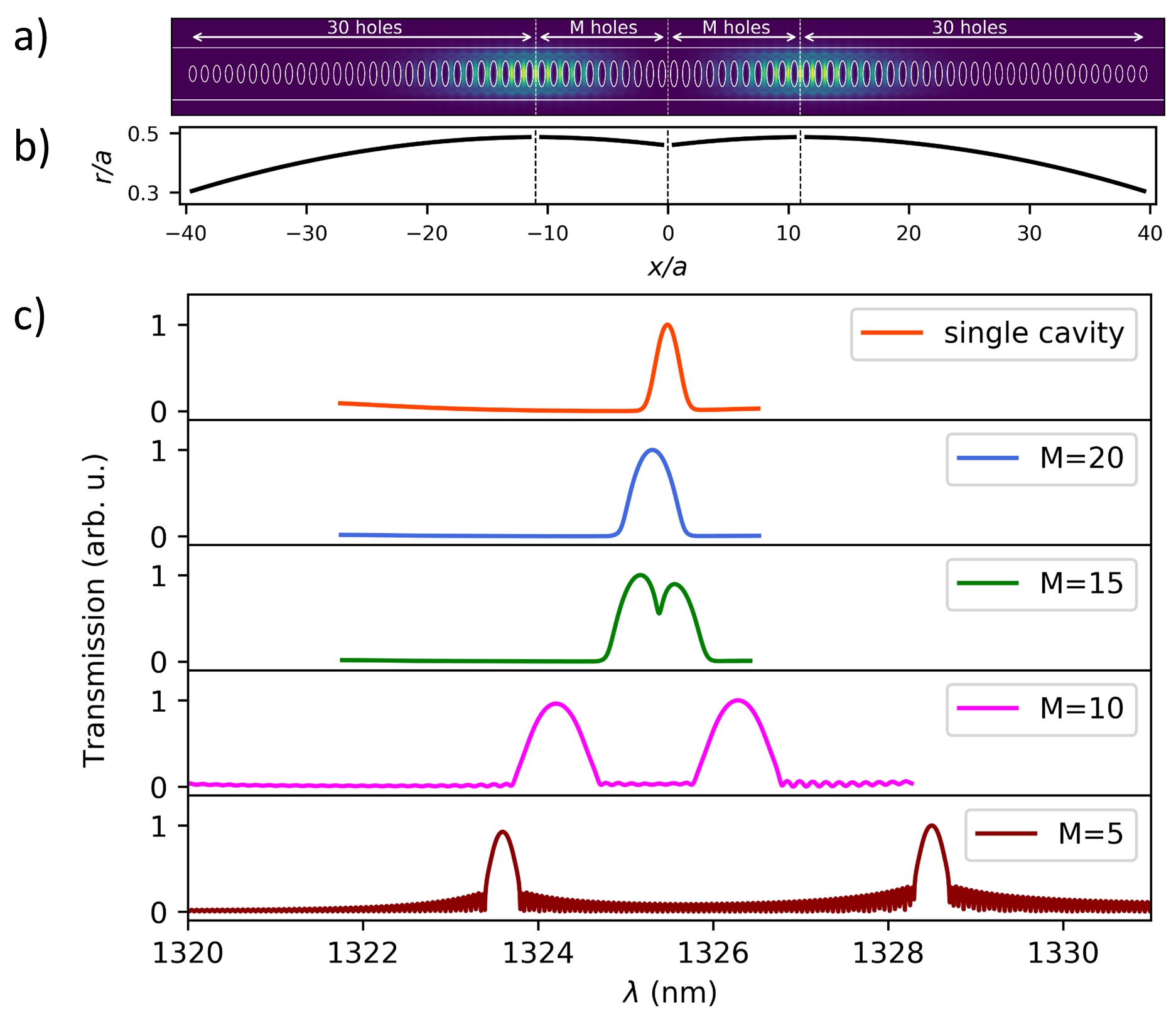}
    \caption{a) Top-view profile of the resonant mode of the photonic molecule. Color bar is the same as in Figure \ref{fig:fig4}. b) Variation of $r$ along the length of the beam. c) Transmission spectrum of a photonic crystal molecule connected via $2M$ holes, in comparison with a single cavity case. 
    }
    \label{fig:fig6}
\end{figure}

Triangular photonic crystal geometry lends itself well toward the 1D coupled cavity array fabrication. To study the coupling between the neighboring cavities, a photonic crystal molecule was designed by stacking two photonic crystal cavities side-by-side with a reduced number of holes in the interior ($M < 30$) as shown in Figure \ref{fig:fig6}a. The major axis radius of the M holes is tapered using the same parabolic variation as the 30 holes with the M largest holes kept and 30-M holes discarded. The coupling between the two cavities causes the resonant mode splitting. We observe in the transmission spectra that as the separation between the two cavities is reduced ($M$ decreases), the coupling strength between the cavities represented in the wavelength separation between the two resonances increases \cite{atlasov2008wavelength}. By integrating emitters into the photonic crystal molecule, the designed structures can be employed for investigation of cQED effects \cite{majumdar2012cavity}. 

\section{Multi-emitter cavity QED in SiC photonic devices}
Strong coupling between light and matter lies at the crux of many applications of optics, including all-optical quantum gates for quantum computing \cite{PhysRevLett.85.2392, PhysRevLett.92.127902} and quantum simulations of strongly correlated condensed matter systems \cite{greentree2006quantum}. While cQED physics has been realized in single quantum dot systems \cite{najer2019gated, RevModPhys.87.347} and single defects in semiconductors \cite{doi:10.1021/acs.nanolett.7b05075}, the integration of multiple quantum emitters into cavities has been suggested as a method of accessing the strong coupling regime for emitters with smaller cavity-emitter coupling terms \cite{radulaski2017nonclassical}. In a multi-emitter system, the $N$ emitters that are coupled individually to the same cavity will effectively behave as a single emitter with emitter-cavity coupling enhanced by a factor of $\sqrt{N}$. 
In order to determine whether this strong coupling effect will occur in the triangular cavity devices described above, the quantum master equation was solved using QuTiP, an open source python package for simulating open quantum systems\cite{johansson}.

The Tavis-Cummings model describes such a multi-emitter system and is given by the Hamiltonian:
$$
H_{TC} = \Omega a^\dagger a +  \sum_{i = 1}^N \left[ \omega_{i} \sigma_i^+ \sigma_i^- + g_i \left( a^\dagger \sigma_i^- + \sigma_i^+ a \right) \right],
$$
where $\Omega$ and $\omega_{i}$ are the cavity frequency and the emitter frequencies, respectively, $a$ and $\sigma_{i}^-$ represent the cavity annihilation operator and the $i$-th emitter lowering operators. Parameters $g_i$ refer to the cavity-emitter coupling rates whose maximal value can be calculated by the equation:
$$g_{max} = \sqrt{\frac{3 \pi c^3 \xi}{2 \tau \omega^2 n^3 V}},$$
and depends on the positioning of the emitter relative to the electric field profile of the cavity by the relationship $g = g_{max}\left| E/E_{max} \right| \cos{\phi}$. $E$ is the electric field intensity at the color center, and $\phi$ is the angle between the orientation of color center dipole and the electric field.
As detailed above, the mode volume of the simulated triangular photonic crystal cavity is $V\sim 2(\lambda/n)^3$, and the values of the optical lifetime $\tau$ and the Debye-Waller factor $\xi$ are dependent upon the choice of the color center (see Table \ref{tab:Table1}). The calculated values of $g/2\pi$ in SiC are given in Table \ref{tab:Table1}.
The loss in the system is simulated via Lindblad dissipators using loss parameters $\kappa$ for cavity and $\gamma$ for each emitter \cite{steck2007quantum}.

\begin{figure}[h!]
    \centering
    \includegraphics{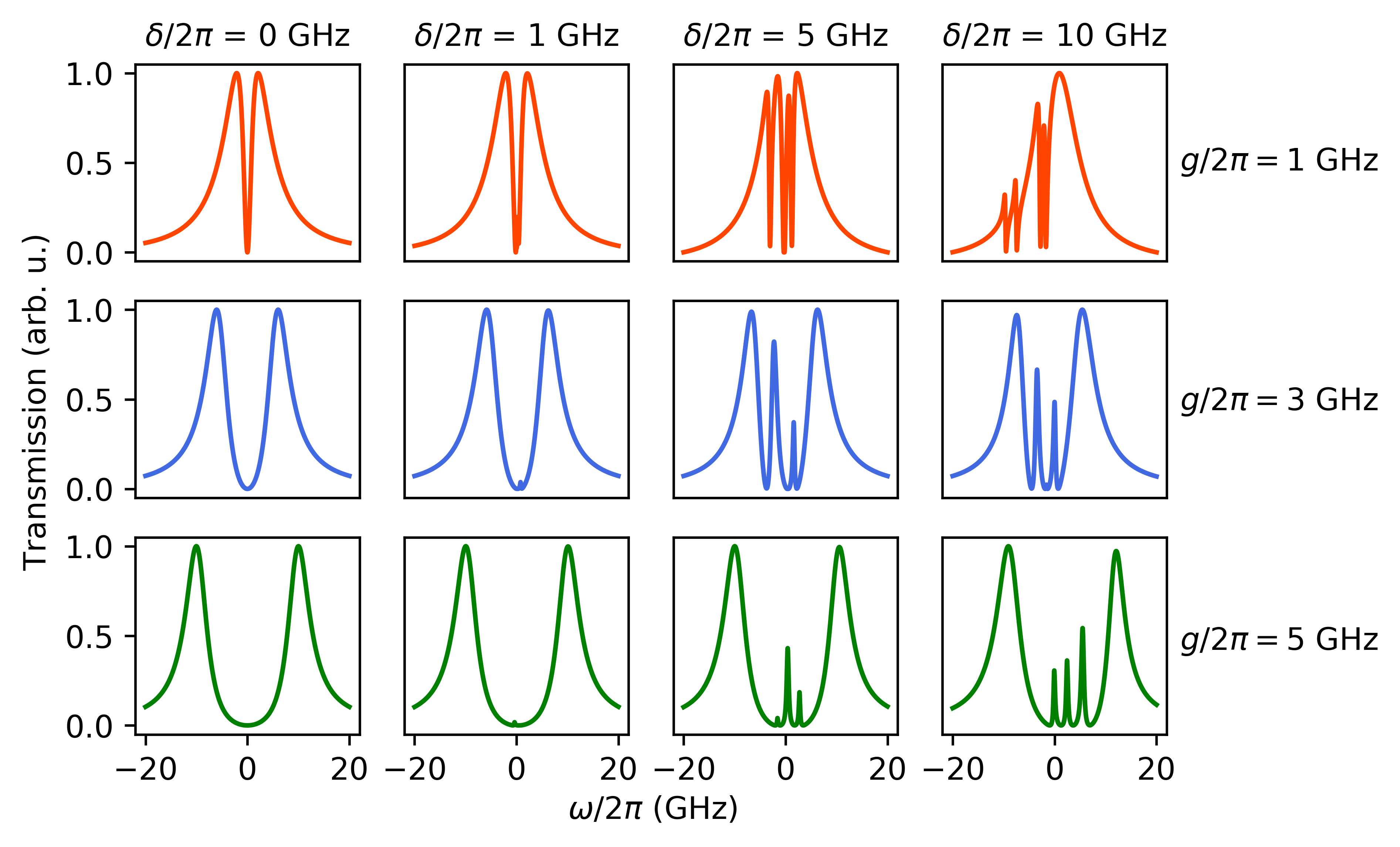}
    \caption{Open quantum system simulations of the Tavis-Cummings model for a single cavity coupled to $N=4$ emitters. The cavities are resonant at $\omega/2\pi = 0$. The emitter frequencies are randomly selected from a Gaussian distribution centered at $\omega = 0$ with a full-width half-maximum of $\delta$. In the case $\delta = 0$, all the emitters are resonant at $\omega = 0$. The emitter frequencies in each panel are randomly generated, independently of those in other panels.
    The cavity and the emitter loss rates are $\kappa/2\pi = 10$ GHz, and $\gamma/2\pi = 0.17$ GHz.
    }
    \label{fig:singlecav}
\end{figure}

A suitable range of $g/2\pi$ values was selected for simulations, and the resulting transmission spectra of a single cavity coupled to $N=4$ emitters are shown in Figure \ref{fig:singlecav}. The emission frequencies of the emitters are randomly generated from a Gaussian distribution centered at the cavity resonance with a standard deviation $\delta/2$, where $\delta$ represents the inhomogeneous broadening in the color center ensemble.
The splitting of the transmission spectrum into two separate polariton peaks separated by $2g\sqrt{N}$ confirms that the system is in the strong cQED coupling regime for both $g/2\pi$ values of 3 GHz and 5 GHz.
Figure \ref{fig:singlecav} shows that the expected cavity-emitter polariton peak splitting persists for values of $\delta \lesssim g\sqrt{N}$. Furthermore, the linewidth of the split emitter-cavity polariton peaks is $\left( \kappa + \gamma \right)/2$, which indicates these systems exist within the cavity-protection domain wherein the emitter ensemble is protected from inhomogeneous broadening caused by the dephasing \cite{zhong2017interfacing}.

\begin{figure}[h!]
    \centering
    \includegraphics{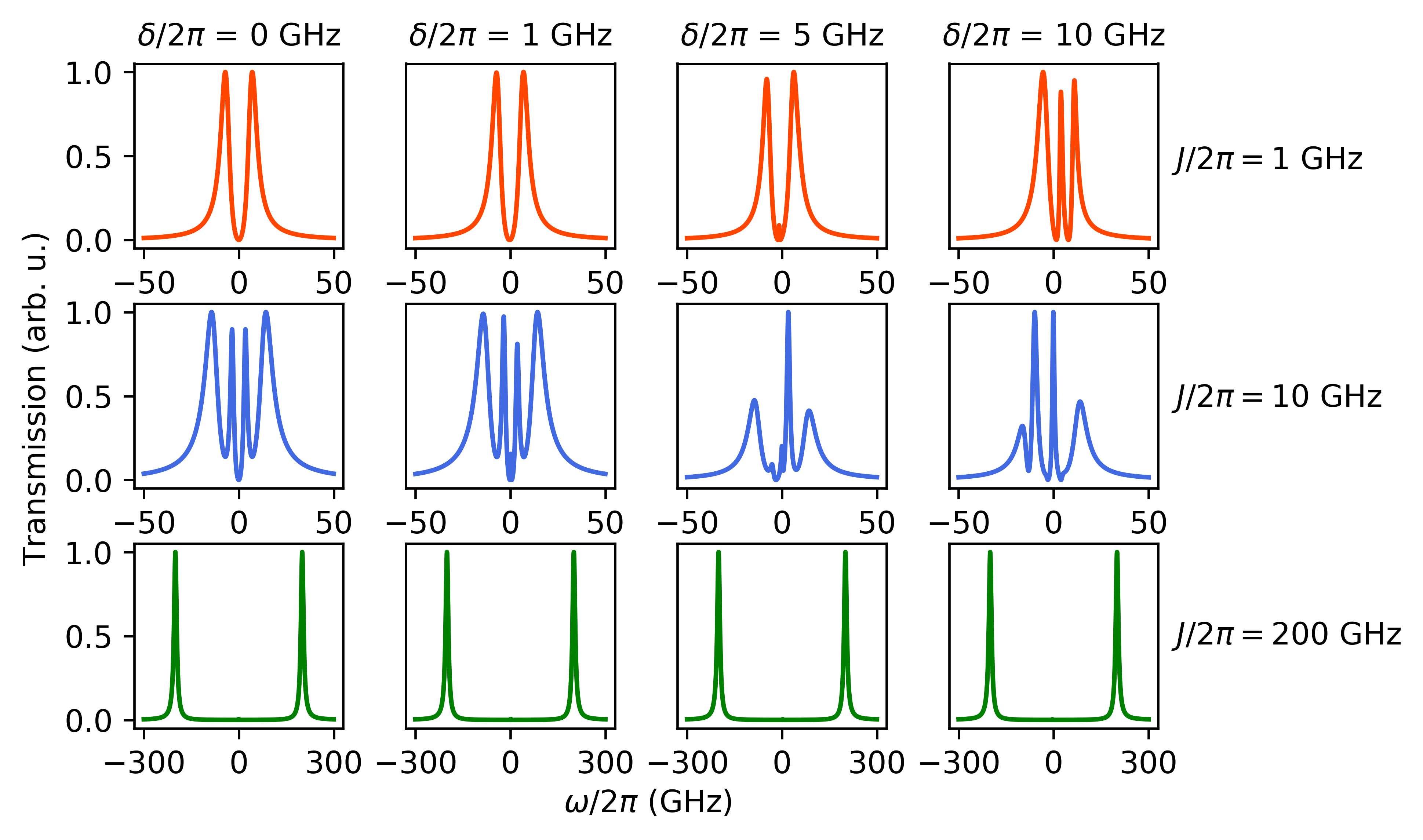}
    \caption{Open quantum system simulations of the Tavis-Cummings-Hubbard model for $n=2$ cavities coupled to $N_i=2$ emitters each at a rate $g/2\pi = 5$ GHz, with the same values for $\kappa$ and $\gamma$ as in Figure \ref{fig:singlecav}. The emitter frequencies in each panel are randomly generated, independently of those in other panels.  The peaks in the all-resonant ($\delta=0$) cases are found at the approximate eigenvalues of the lossless strong coupling case: $\omega = \pm g \sqrt{2} \pm J $ for the $g > J$ case (orange), and $\omega = \pm 2g^2/J, \pm 2g^2/J + J$ for the $g < J$ cases (blue and green). The linewidth of these peaks in the case $\delta < J$ is given by $\left( \kappa + \gamma_i \right)/2$ which is indicative of the cavity protection regime wherein the emitter ensemble is protected from the inhomogeneous broadening by the cavity. The intermediate peaks correspond to the subradiant states and are present even in the $J/2\pi = 200$ GHz simulations (green) though, due to scale, they are not easily visible in the plots.}
    \label{fig:doublecavity}

\end{figure}

Photonic quantum simulation proposals \cite{Plenioreview2008} rely on the use of coupled cavity arrays in the strong coupling regime of cQED. Here, we explore the smallest such system of $n = 2$ coupled nanocavities, as modeled in Figure \ref{fig:doublecavity}, and in relation to the designed photonic crystal molecule in Figure \ref{fig:fig6}. In order to explore the physics of such a system, the Tavis-Cummings model can be expanded upon by adding a cavity coupling term to form the Tavis-Cummings-Hubbard (TCH) model:
$$H_{TCH} = \sum_{i = 1}^n \left\{ \Omega_{i} a_{i}^\dagger a_{i} + \sum_{j = 1}^{N_i} \left[ \omega_{i,j} \sigma_{i,j}^+ \sigma_{i,j}^- + g_{i,j}\left( a_i^\dagger \sigma_{i,j}^- + \sigma_{i,j}^+ a_i \right) \right]  \right\}  - \sum_{i = 1}^{n-1} J_{i, i+1} \left(a_i^\dagger a_{i+1} + a_{i+1}^\dagger a_{i} \right)
$$
where $n$ is the number of cavities, $\omega_{c,i}$, $N_{i}$ and $a_{i}$ are the resonant frequency of, the number of emitters in, and the annihilation operator of the $i$-th cavity. The resonant frequency of the $j$-th emitter in the $i$-th cavity is $\omega_{i,j}$ and it has the annihilation operator $\sigma_{i,j}^-$. The coupling rate between neighboring cavities is given by $J_{i, i+1}$.

The transmission spectra of $n=2$ cavities with $N_i=2$ emitters coupled to each cavity for $g/2\pi = 5$ GHz were simulated in QuTiP with and without broadening in emitter frequencies and the results are shown in Figure \ref{fig:doublecavity}. We extracted values of the photon hopping rate $J$ by fitting a two-peak Lorentzian to the plots in Figure \ref{fig:fig6}; the splitting of polariton peaks in strongly coupled systems is directly proportional to the coupling strength. The $J/2\pi = 1$ GHz and $J/2\pi = 200$ GHz photonic crystal molecule cQED plots exhibit the same cavity protection linewidth signature and peak splitting as the single cavity simulations. The structure of the $J/2\pi = 10$ GHz set of simulations merits some closer discussion; the eigenvalues of the lossless TCH model when $J < g$ in a fully resonant case are $\omega = \pm 2g^2/J$ and $\omega = \pm J + 2g^2/2$, as seen in the $\delta/2\pi = 0$ GHz, $J/2\pi = 10$ GHz plot in Figure \ref{fig:doublecavity}. This same set of peaks is maintained for the values of $\delta \lesssim g\sqrt{N_i}$, which means that light-matter hybridized states continue even with the introduction of modest, but realistic, inhomogeneous broadening into the system. These results demonstrate that the designed triangular cross-section devices are strong candidates for quantum information and simulation platforms. In addition, the inhomogeneously broadened emitter systems exhibit subradiant peaks, which have been studied in single-cavity systems and found to facilitate coherence conducive to high-purity quantum light generation \cite{radulaski2017photon}.

\section{Discussion}

Our modeling results provide insights into color center integration with triangular SiC photonic devices, as well as quantum optical phenomena in cavity QED systems. This geometry is of great interest because it overcomes the limitations of color center platforms requiring implementation in bulk substrates.

Triangular-cross-section waveguides are found to efficiently guide the light emission from color centers embedded at optimal depths. The waveguides support propagation of both TE and TM polarized modes, thus serving a variety of color center crystalline orientations. Fabrication imperfections in the curvature of the apex affect only the modes present in the lower half of the triangular cross-section. Similar designs fabricated in diamond have successfully combined triangular waveguide geometry with suitable grating couplers for efficient collection of light into a high numerical aperture objective, as well as tapered down for coupling to an optical fiber \cite{burek2017fiber}. They can also be designed for integration with the emerging superconducting nanowire single-photon detectors \cite{martini2019single}.

The modeled triangular nanobeam photonic crystal cavities have high $Q$/$V$ ratio with promising applications in high-speed and indistinguishable photon generation. For the same triangular geometry, altering of the lattice constant is found to tune the resonant wavelength without reduction in the quality factor, meaning that many different high-performing devices can be simultaneously fabricated on the same chip. Interestingly, in literature on diamond and SiC nanobeam devices, the fabricated quality factors fell 2-3 orders of magnitude short of their theoretical predictions \cite{burek2014high, song2018high}. The reduced performance can be attributed to fabrication errors such as mask alignment imperfections \cite{song2018high}. This indicates that additional efforts in nanofabrication process development need to be invested to realize the full application potential of these designs. Purcell enhancement of the zero-phonon line emission would play a crucial role in realizing quantum systems that require two-photon interference and quantum entanglement, such as quantum repeaters \cite{azuma2015all} and cluster-entangled states \cite{raussendorf2001one,economou2016spin}. The photonic crystal cavities modeled in this work potentially provides a tie-in between high $Q$/$V$ designs and platforms that host color centers, a necessary step for color center-based quantum information technologies in SiC.

The open quantum system modeling of a single nanocavity and a photonic crystal molecule integrating non-identical color centers confirmed that cavity-protection effects can support polariton physics even in the presence of realistic inhomogeneous broadening in SiC structures. Extensions to larger 1D coupled-cavity-arrays can provide a simulation testbed for proposals of strongly-correlated systems, such as the 1D Mott insulator to superfluid transition \cite{Hartmann2008}, and could be extended to waveguide QED simulations of the many-body localization phase \cite{fayard2021manybody}.

Our simulations of triangular cross-section waveguides and cavities in 4H-SiC demonstrate that this geometry can achieve similar figures of merit (e.g. \textit{Q} $\sim 10^6$) as other state-of-the-art geometries in both 4H- and 3C-SiC polytype. Since, the thickness and width of these devices are related through the etch angle, global scaling is possible without affecting the figures of merit as much. The benefit to our approach is that 4H-SiC can host color centers without damaging the lattice unlike 3C-SiC and the geometry is highly scalable using angle RIE etching unlike some of the more complex 2D photonic crystal cavities. In fact, triangular cross-section nanophotonics have already been successfully fabricated in diamond; the work already performed in diamond has illuminated some shortfalls in nanofabrication that need to be overcome before we can fully realize our simulated results. The prior experiments and our work in this paper pave the way toward high quality, color center integrated, highly scalable nanophotonic devices in 4H-SiC.

\section{Acknowledgments}

This work is supported by the National Science Foundation (CAREER-2047564) and the UC Davis Summer GSR Award for Engineering or Computer-related Applications and Methods. The authors thank Richard Scalettar and Jesse Patton for fruitful discussions on Tavis-Cummings-Hubbard physics.

\bibliographystyle{unsrt}
\bibliography{biblio.bib}
\end{document}